\documentclass[superscriptaddress,groupedaddress,nofootnoteinbib,11pt]{article}

\usepackage{jcappub}
\usepackage{bm}
\usepackage{verbatim}

\usepackage{epsf}
\usepackage{graphicx,epsfig}
\usepackage{amsfonts}
\usepackage{amssymb}
\usepackage{xcolor}

\definecolor{mine}{rgb}{0.2,0.1,0.7}
\definecolor{bb}{rgb}{0.3, 0.5, 1}
\definecolor{bg}{rgb}{0.1, 0.1, 0.5}

\def\half{\frac12}

\def\K{{\cal K}}

\def\g{\gamma}
\def\P{\tilde \pi}
\def\L{\Lambda}

\def\A{A}

\def\T{T}

\def\mpl{M_{\rm Pl}}
\def\e{{\epsilon}}

\def\d{\mathrm{d}}

\def\K{{\cal K}}

\def\L*{{\cal L}_*}
\def\L{\mathcal{L}}
\def\({\left(}
\def\){\right)}
\def\ie{{\it i.e. }}
\def\nn{\nonumber}
\def\mn{_{\mu \nu}}
\def\stu{St\"uckelberg }
\def\p{\partial}
\def\mupn{^\mu_{\ \nu}}
\def\<{\langle}
\def\>{\rangle}
\def\Ein{\hat{\mathcal{E}}}

\def\e{\varepsilon}

\newcommand{\bea}{\begin{eqnarray}}
\newcommand{\eea}{\end{eqnarray}}
\newcommand\be{\begin{equation}}
\newcommand\ee{\end{equation}}
\newcommand\beq{\begin{equation}}
\newcommand\eeq{\end{equation}}

\def\ba{\begin{eqnarray}}
\def\ea{\end{eqnarray}}

\newcommand{\refeq}[1]{(\ref{#1})}

\begin{document}

\title{Massive Gravity on de Sitter and Unique Candidate for Partially Massless Gravity}

\author[1]{Claudia de Rham}
\author[2,3]{and S\'ebastien Renaux-Petel}
\affiliation[1]{Department of Physics, Case Western Reserve University, 10900 Euclid Ave, Cleveland, OH 44106, USA}
\affiliation[2]{Centre for Theoretical Cosmology,
Department of Applied Mathematics and Theoretical Physics,
University of Cambridge, Cambridge CB3 0WA, UK }
\affiliation[3]{UPMC Univ Paris 06, CNRS, Institut Lagrange de Paris, Laboratoire de Physique Th\'eorique et Hautes Energies, UMR 7589, 4 place Jussieu, 75252 Paris Cedex 05, France}
\vskip 4pt

\date{\today}


\abstract{
We derive the decoupling limit of Massive Gravity on de Sitter in an arbitrary number of space-time dimensions $d$. By embedding $d$-dimensional de Sitter
into $d+1$-dimensional Minkowski, we extract the physical helicity-1 and helicity-0 polarizations of the graviton. The resulting decoupling theory is similar to that obtained around Minkowski. We take great care at exploring the partially massless limit and define the unique fully non-linear candidate theory that is free of the helicity-0 mode in the decoupling limit, and which therefore propagates only four degrees of freedom in four dimensions. In the latter situation, we show that a new Vainshtein mechanism is at work in the limit $m^2\to 2 H^2$ which decouples the helicity-0 mode when the parameters are different from that of partially massless gravity. As a result,  there is no discontinuity between massive gravity and its partially massless limit, just in the same way as there is no discontinuity in the massless limit of massive gravity. The usual bounds on the graviton mass could therefore equivalently well be interpreted as bounds on $m^2-2H^2$.
When dealing with the exact partially massless parameters, on the other hand, the symmetry at $m^2=2H^2$ imposes a specific constraint on matter. As a result the helicity-0 mode decouples without even the need of any Vainshtein mechanism.}

\maketitle


\section{Introduction}

As the search for Gravitational Waves is about to enter its golden age, with advanced LIGO \cite{Abramovici:1992ah}, VIRGO \cite{Caron:1997hu}, GEO 600 \cite{Willke:2002bs}, TAMA 300 \cite{Tsubono:1994sg}, as well as many other interferometers and CMB probes such as the South Pole Telescope \cite{Ruhl:2004kv} or Planck \cite{Bouchet:2007zz} for the detection of primordial Gravitational Waves, the time is right to reflect on what can be expected from a purely theoretical perspective. Even though General Relativity (GR) predicts the propagation of two polarizations, which have been indirectly tested with an unchallenged accuracy via binary pulsars timing delay, the presence of additional polarizations might not necessarily be ruled out, neither from a theoretical nor from an observational point of view. If unprotected by a symmetry, a four-dimensional rank-two symmetric  tensor field can in principle propagate up to six polarizations\footnote{See Ref. \cite{VanNieuwenhuizen:1973fi} for more general tensor theories that could propagate more degrees of freedom.}, namely two helicity-2 modes, two helicity-1 modes and finally two helicity-0 modes. Whilst one of these helicity-0 modes typically comes hands in hands with a ghost-like instability\footnote{There could of course be more propagating degrees of freedom when including additional scalar or vector fields, but we focus here only on the fundamental degrees of freedom of the graviton itself.}, the other five, on the other hand could be perfectly healthy. Massive gravity, be it a soft mass or a hard mass, is one  of the most natural examples of a theory that propagates five polarizations for the graviton (such as in the DGP (Dvali-Gabadadze-Porrati) model \cite{Dvali:2000hr}, Lorentz violating massive gravity \cite{Dubovsky:2004sg}, massive gravity \cite{Fierz:1939ix,deRham:2010ik,deRham:2010kj} {\it etc}). H$\breve{\rm o}$rava-Lifschitz is another example which could contain up to two and a half degrees of freedom, \cite{Horava:2009uw,Blas:2009yd,Blas:2009qj}.
Actually the list is long since any modification of gravity necessarily comes with new degrees of freedom, and especially scalar ones. Interestingly, it turns out that phenomenological bounds on modified gravity, be it for instance on the speed of the gravitational waves or on the graviton mass, are mainly driven by bounds on the presence of these new scalar degrees of freedom, \cite{Will:2005va}.

With this in mind, we will therefore explore in this manuscript whether there could be a consistent modification of gravity which does not propagate any helicity-0 mode. As mentioned previously, when properly taken care of, massive gravity defined on flat space propagates not six but only five degrees of freedom, including one helicity-0 mode.
In order to project out this last helicity-0 mode, one can attempt a relatively benign modification and consider instead that theory around a  maximally symmetric reference metric, \cite{Hassan:2011tf}.  At the linearized level, the presence of a constant curvature for the reference metric only affects the helicity-0 mode and one can tune the graviton mass with respect to this curvature in such a way that the helicity-0 mode entirely disappears at the linearized level. This remarkable feature of massive gravity on de Sitter was first pointed out by S.~Deser and A.~Waldron, \cite{Deser:2001pe,Deser:2001us,Deser:2001wx,Deser:2001xr,Zinoviev:2001dt,Deser:2004ji,Deser:2006zx,Gabadadze:2008uc}
 and referred to as ``Partially Massless" (PM) gravity. In this paper we will explore this specific theory further by following the interactions present in the full theory and by providing the unique candidate for the fully non-linear PM theory of gravity.

We start our analysis by studying massive gravity on de Sitter at the non-linear level and base ourselves on the fully non-linear theory of massive gravity proposed in \cite{deRham:2010kj}, which has been shown to be free of ghost as much around Minkowski  (see Refs.~\cite{deRham:2010ik,deRham:2010kj,Hassan:2011hr,Hassan:2011ea,deRham:2011rn,deRham:2011qq,Hassan:2012qv,Mirbabayi:2011aa})  than around any reference metric, be it dynamical or not \cite{Hassan:2011tf,Hassan:2011zd}. In order to focus on the relevant interactions for the helicity-0 mode, we work in a so-called decoupling limit where the helicity-0 mode (and the helicity-1 modes) may be treated non-linearly while the helicity-2 modes are kept linear. This decoupling limit is merely a useful trick to identify the relevant physics up to a given energy scale and to disentangle the relevant interactions of the helicity-0 mode from the standard (and well understood) complications and non-linearities of General Relativity. We take this decoupling limit by sending simultaneously $\mpl \to \infty$ and $m, H\to 0$ (where $R=d(d-1)H^2$ is the scalar curvature of the de Sitter metric in $d$ dimensions) while keeping the scales $\Lambda^3=\mpl m^2$ and $m/H$ fixed\footnote{When the reference metric is Anti de Sitter, the decoupling limit may be taken differently and the curvature of the reference metric does not need to vanish in the decoupling limit. These aspects will be exposed soon in \cite{AdS}.}.

When deriving the decoupling limit of massive gravity on a flat reference metric, the expression of the metric in terms of the helicity-2, -1 and -0 modes is relatively straightforward, and can be achieved by use of the \stu trick, \cite{ArkaniHamed:2002sp}. However as soon as one departs from Minkowski, the mixing with the reference curvature makes the use of the \stu trick much more subtle. In this paper we bypass any ambiguity by embedding the maximally symmetric reference metric into a flat higher-dimensional space-time in which the definition of the \stu fields takes the standard form. By then projecting their expression back onto the lower-dimensional manifold, we obtain a fully non-linear expression for the covariantized reference metric in terms of the properly identified \stu fields. This method is complementary to the one used in \cite{Mirbabayi:2011aa} and can in principle be used for a large variety of reference metrics, although we only focus on the de Sitter one in this manuscript.

Using this logic, we successfully derive the decoupling limit for massive gravity on de Sitter in all generality and in an arbitrary number of dimensions. We find a decoupling limit which is qualitatively very similar to that on Minkowski, and manifests the presence of a series of Galileon terms. When exploring this decoupling limit in more depth, we unveil the existence of a fully non-linear candidate for $d$-dimensional PM gravity for very specific choices of the parameters present in the Ghost-free theory of gravity established in \cite{deRham:2010ik}, namely $\alpha_3=-\frac13 \frac{d-1}{d-2}$ and $\alpha_n=-\frac{1}{n} \alpha_{n-1}\, \, {\rm for} \, \, n \geq 4$. If a fully non-linear PM theory exists, then this method uniquely fixes its interactions to all order within and beyond the decoupling limit. This analysis does not allow us to establish whether the absence of the helicity-0 mode persists beyond the decoupling limit, but if PM has any chance of existing, it has to be the theory presented here.

We also point out an important distinction between this result and the decoupling limit of the ``minimal model". In the minimal model on flat space, all the interactions for the helicity-0 mode disappear in the decoupling limit. However the kinetic term remains present and the four-dimensional minimal model propagates five degrees of freedom. One simply needs to probe energy scales above $\Lambda$ to excite these interactions. In the PM model on the other, the helicity-0 mode also looses its kinetic terms, so there is no sign of the helicity-0 excitation, both at the linear level and at all orders in the decoupling limit.

Finally, considering massive gravity on de Sitter rather than Minkowski might seem at first as defeating some of the objectives of considering a Lorentz invariant theory of massive gravity since Lorentz invariance is broken on de Sitter. However the amount of symmetry on de Sitter remains the same as on Minkowski so one still keeps control over a larger symmetry group in massive gravity on de Sitter compared to a generic Lorentz violating theory of gravity.

The rest of this paper is organized as follows: we review the recent developments in Massive Gravity and BiGravity in section \ref{sec:MG} and highlight the prescription to derive the standard decoupling limit around Minkowski. We then present the higher-dimensional framework we use to covariantize the reference metric in section \ref{sec:5d} and properly introduce the \stu fields on de Sitter. We leave some of the technical details for appendix \ref{sec:Coord-transformation}. Using this technique, we derive the linearized action in section \ref{sec:Linearized} and highlight the key features of the theory. In particular we comment on the Higuchi bound, on the existence of a  vDVZ discontinuity and on the PM special case. We then derive the full decoupling limit of the theory in section \ref{sec:DL} and appendix \ref{sec:appendix dS} before defining and exploring the PM theory of gravity in section \ref{sec:PM}. We finally conclude and present some open questions in section \ref{sec:Outlook}.

\section{Review of Massive Gravity}
\label{sec:MG}

\subsection{Massive Gravity around Minkowski}

The Fierz-Pauli action was the first attempt to construct a theory of massive gravity in  an arbitrary number of dimensions $d>2$, \cite{Fierz:1939ix}
\ba
\label{FP}
\L_{\rm FP}= -\frac {\mpl^{d-2}}4 \tilde h^{\mu\nu}\Ein^{\alpha\beta}\mn \tilde h_{\alpha\beta} - \frac {\mpl^{d-2} m^2}8  \(\tilde h_{\mu\nu}\tilde h^{\mu\nu}-\tilde h^\mu_\mu \tilde h^\nu_\nu\)\,,
\ea
which is valid at first order in perturbations around flat space-time, $g\mn= \eta\mn+\tilde h\mn= \eta\mn+ h\mn/\mpl^{(d-2)/2}$, $\tilde h\mupn=\eta^{\mu\alpha}\tilde h_{\alpha \nu}$ and $\Ein$ is the Lichnerowicz operator,
\ba
\Ein^{\alpha\beta}\mn \tilde h_{\alpha\beta}=-\frac 12 \(\Box \tilde h\mn-2\p_{(\mu}\p_\alpha  \tilde h^\alpha_{\nu)}+\p_\mu\p_\nu \tilde h-\eta\mn (\Box \tilde h-\p_\alpha\p_\beta \tilde h^{\alpha\beta})\)\,,
\ea
where we use the convention $(A,B)=\frac 12 (AB+BA)$.
As is well known, the non-linearities beyond this linearized Fierz-Pauli action are essential for the consistency of the theory and in order to avoid any discontinuity in the massless limit $m\to 0$, \cite{vDVZ,Vainshtein:1972sx}. Non-linearities however come with their share of troubles as they usually lead to the excitation of an additional mode known as the Boulware-Deser (BD) ghost, \cite{Boulware:1973my}. For the theory to remain consistent at the non-linear level, the mass term should be implemented with a specific set of non-linear interactions, \cite{deRham:2010ik}, leading to the fully non-linear action, \cite{deRham:2010kj}
\ba
\label{MG_lagrangian}
\mathcal{L}_{MG}=\frac{\mpl^{d-2}}{2}\sqrt{-g}\left(R-\frac{m^2}{4}\mathcal{U}(g)\right)\,,
\ea
where the most general potential $\mathcal{U}$ is
\ba
\label{pot}
\mathcal{U}(g)=-4 \sum_{n=2}^d \alpha_n \L_{\rm der}^{(n)}(\K)\,.
\ea
For definiteness we will always choose $\alpha_2=1$ while the other coefficients $\alpha_n$ are {\it a priori} arbitrary.
The tensor $\K\mn$ is defined as $\mathcal{K}^{\mu}_{\nu}(g)=\delta^{\mu}_{\nu}-\sqrt{g^{\mu\alpha}\eta_{\alpha\nu}}$
and
\ba
\L_{\rm der}^{(n)} (\K)=-\sum_{m=1}^{n}(-1)^m \frac{(n-1)!}{(n-m)!}   \langle \K^m \rangle  \L_{\rm der}^{(n-m)} (\K)\,,
\ea
where $\L_{\rm der}^{(0)} (\K)=1$ and $\L_{\rm der}^{(1)} (\K)=  \langle \K \rangle$, or more explicitly,
\ba
\label{U2}
\L_{\rm der}^{(2)}(\K)&=&  \langle \mathcal{K}\rangle^2- \langle \mathcal{K}^2\rangle,\\
\label{U3}
\L_{\rm der}^{(3)}(\K)&=&  \langle \mathcal{K}\rangle^3-3 \langle \mathcal{K}\rangle  \langle\mathcal{K}^2\rangle+2 \langle\mathcal{K}^3\rangle,\\
\label{U4}
\L_{\rm der}^{(4)}(\K)&=&  \langle\mathcal{K}\rangle^4-6 \langle\mathcal{K}^2\rangle \langle\mathcal{K}\rangle^2+8 \langle\mathcal{K}^3\rangle \langle\mathcal{K}\rangle+3 \langle\mathcal{K}^2\rangle^2-6 \langle\mathcal{K}^4\rangle\,, \\
{\rm etc}\ldots \ \ && \nn
\ea
where $\langle \ldots \rangle$ represents the trace of a tensor with respect to the metric $g\mn$, whilst in the following of the paper $\left[\cdots\right]$ will represent the trace with respect to the reference metric, be it Minkowski or de Sitter depending on the specific case.

The absence of ghost for this theory has been shown in the decoupling limit in \cite{deRham:2010ik}, whilst the argument beyond the decoupling limit was provided in \cite{deRham:2010kj} and shown to work to all orders in \cite{Hassan:2011hr,Hassan:2011ea}. For complementary arguments in the \stu and helicity languages see Refs.~\cite{deRham:2011rn,deRham:2011qq,Hassan:2012qv}. Furthermore, the linearized theory has also been recently derived around arbitrary backgrounds (when working in the local inertial frame) in Ref.~\cite{Mirbabayi:2011aa}. This analysis shows that the absence of ghost in the decoupling limit also ensures the consistency of the theory fully non-linearly.

As it will be apparent later, it is also useful to express the Lagrangians $\L_{\rm der}^{(n)}$ in terms of the fully antisymmetric Levi-Cevita tensor\footnote{We defined the Levi-Cevita tensor as $\mathcal{E}^{\alpha_1 \cdots \alpha_d}=\sum_{\sigma}(-1)^\sigma\delta^{\alpha_1}_{\sigma(0)}\cdots \delta^{\alpha_d}_{\sigma(d-1)}$, where the sum is over all the possible permutations of $\{0,\cdots, d-1\}$.},

\ba
\L_{\rm der}^{(n)} (\K)=-\frac{1}{(d-n)!}  \mathcal{E}^{\alpha_1 \cdots \alpha_d} \mathcal{E}_{\beta_1 \cdots\beta_n \alpha_{n+1} \cdots \alpha_d}\
\K_{\alpha_1}^{\beta_1}\cdots \K_{\alpha_n}^{\beta_n}\,,
\ea
where indices are raised and lowered using the metric $g\mn$.
As is clear from this formulation, one can only construct such Lagrangians up to $n\le d$ before running out of indices. It will also be useful to define the Galileon Lagrangians
\ba
\L_{\rm Gal}^{(n)} = (\p \P)^2 \L^{(n-2)}_{\rm der}(\tilde \Pi)\,.
\ea
We stress that, here and in the following, the traces are taken with respect to the reference metric in $\L_{\rm der}^{(n)}(\tilde \Pi)$.

\subsection{Helicity-0 and -1 modes}
\label{sec:helicity0_Minkowski}

As is clear from the Fierz-Pauli action \eqref{FP}, the mass term explicitly breaks covariance (or its linear diffeomorphism realization). To restore it, one can resort to the well-known \stu trick, first introduced to by \stu in 1938 to restore gauge-invariance in electromagnetism and formalized among others by Glashow in 1962, but which works equivalently well for spin-2 fields, \cite{Siegel:1993sk,ArkaniHamed:2002sp,Creminelli:2005qk}. The key of the \stu approach is to introduce $d$ fields $\phi^a$, which transform as scalars under coordinate transformation, so as to adapt for the fixed metric $\eta\mn$. More precisely, the fixed metric $\eta\mn$ is promoted to a tensor field
\ba
\tilde \eta\mn=\eta_{ab} \p_\mu \phi^a\p_\nu \phi^b\,,
\ea
such that the quantity $g^{\mu\alpha}\tilde \eta_{\alpha\nu}$ transforms as a tensor, and the potential $U(g, \tilde \eta)$ is a scalar. For clarity reasons, we will stick to four dimensions for the rest of this section, but exactly the same arguments are valid in any number of dimensions.

By writing the potential in this explicitly covariant form, we introduce gauge invariance (which allows us to set a gauge for $h_{\mu \nu}$ and show that it contains only 2 dynamical degrees of freedom) but by the same token we also introduce 4 degrees of freedom $\phi^a$. However with the specific potentials introduced in (\ref{U2}-\ref{U4}), only three out of these four \stu fields end up being dynamical \cite{deRham:2010ik,deRham:2010kj,deRham:2011rn,Mirbabayi:2011aa}. This is most easily seen in the decoupling limit, where the interactions between the usual helicity-2 modes present in GR and the additional helicity-1 and -0 modes arising in massive gravity take a simpler form.

The decoupling limit of massive gravity is taken by sending the Planck scale $\mpl \to \infty$ so as to reduce any non-linearities that usually arise in General Relativity, while keeping the scale $\Lambda=(\mpl m^2)^{1/3}$ fixed so as to keep interactions arising at or below that scale. In this decoupling limit, all fields are living around the flat Minkowski metric, $g\mn=\eta\mn+\tilde h\mn= \eta\mn+h\mn/\mpl \to \eta\mn$ (where the canonically normalized field $h$ is kept constant in that limit). So in addition to the local four-dimensional diffeomorphism invariance ($h\mn \to h\mn +\p_{(\mu}\zeta_{\nu)}$ with $\phi^a \to \phi^a$), and to the internal Global Lorentz Invariance with Lorentz transformation matrix $\tilde \Lambda^a_{\ b}$,
\ba
x^\mu \to x^\mu\,,\hspace{20pt}
\tilde h\mn \to \tilde h\mn \hspace{10pt}{\rm and}\hspace{10pt}
\phi^a \to \tilde \Lambda^a_{\ b} \phi^b\,,
\ea
we recover one additional accidental global symmetry: space-Time Global Lorentz Invariance with Lorentz transformation matrix $\Lambda\mupn$,
\ba
x^\mu \to \Lambda\mupn x^\nu\,,\hspace{20pt}
\tilde h\mn \to \Lambda^\alpha_{\ \mu}\Lambda^\beta_{\ \nu} \tilde h_{\alpha \beta}\hspace{10pt}{\rm and}\hspace{10pt}
\phi^a \to \phi^a\,.
\ea
The Space-Time and Internal Lorentz symmetries are independent, however if we identify both groups $\tilde \Lambda = \Lambda\equiv \bar \Lambda$ and work in the representation of the single group with
\ba
x^\mu \to \bar \Lambda \mupn x^\nu\,,\hspace{20pt}
\tilde h\mn \to \bar \Lambda^\alpha_{\ \mu}\bar \Lambda^\beta_{\ \nu} \tilde h_{\alpha \beta}\hspace{10pt}{\rm and}\hspace{10pt}
\phi^a \to \bar \Lambda^a_{\ b} \phi^b\,,
\ea
then the combined \stu fields $\{\phi^a\}$ behave as a vector under this global symmetry. It then makes sense to use a standard Scalar-Vector-Tensor decomposition of the degrees of freedom, and write $\phi^a= x^a -\tilde A^a -\p^a \tilde \pi$, where now $\tilde A^a$ behaves as a vector field (and bears two degrees of freedom) and $\tilde \pi$ behaves as a scalar field. In this limit where gravity decouples and every field lives on flat space-time, the vector field $\tilde A^a$ then captures the physics of the helicity-1 modes whilst the scalar field $\tilde \pi$ captures the physics of the helicity-0 mode. We emphasize however that this split  $\phi^a= x^a -\tilde A^a -\p^a \tilde \pi$ is only expressed in terms of the helicity-0 and -1 modes when working in the decoupling limit, and that beyond the decoupling limit (or as we shall see below, around a different background), this split does not successfully exhibit the full helicity-0 and -1 degrees of freedom \cite{deRham:2011rn}.

\subsection{Decoupling limit}

To work out the relevant interactions that arise at the energy scale $\Lambda=(\mpl m^2)^{1/3}$, one should first canonically normalize the different degrees of freedom. As is known from General Relativity, the canonically normalized helicity-2 mode is $h\mn=\mpl \tilde h\mn$, and a linear analysis shows that the correctly  canonically normalized helicity-1 and -0 modes are given by $A^a=\mpl m \tilde A^a$ and $\pi = (\mpl m^2) \tilde \pi$. Non-linearities in the helicity -1 and -2 modes therefore disappear in the decoupling limit where $\mpl \to \infty $ and $\mpl m \to \infty$, whilst non-linearities for the helicity-0 mode could be relevant since $\mpl m^2=\Lambda^3\to$ Const. In what follows we omit any contributions from the helicity-1 modes as they completely decouple in this limit.  The tensor $\mathcal{K}$ is built precisely so as to reduce to $\Pi\mn=\p_\mu \p_\nu \pi$ in the decoupling limit around Minkowski, and so to leading order in $h_{\mu \nu}$ we see that the potential are just total derivatives, $\L_{\rm der}^{(2)}(\Pi)= [\Pi]^2 - [\Pi^2]=$ total derivative and similarly for $\L_{\rm der}^{(3)}$ and $\L_{\rm der}^{(4)}$. Ignoring the vector field, the next to leading contribution from the potential is  then given by
\ba
m^2 \mpl^2 \sqrt{-g}\,\mathcal{U}(g,\phi)_{\rm dec} =  \Lambda^3 h\mn X^{\mu\nu}
\ea
with
\ba
X^{\mu\nu}=\mpl \(\frac{\delta}{\delta h\mn}\sqrt{-g}\, \mathcal{U}(g,\phi)\)\Big|_{h\mn=0}\,.
\ea
In terms of the helicity-0 mode, this tensor $X\mn$ can be shown to be \cite{deRham:2010ik,deRham:2010kj}
\ba
X\mn=-4 \left[ 2 X^{(1)}\mn+(1+3 \alpha_3)X^{(2)}\mn+ (\alpha_3 + 4 \alpha_4)X^{(3)}\mn \right]
\ea
where the tensor $X^{(n)}\mn$ is defined as
 \ba
 \label{Xmn}
 X^{(n)}\mn=\sum_{m=0}^n (-1)^m \frac{n!}{2(n-m)!}\tilde \Pi^m\mn\L^{(n-m)}_{\rm der}(\tilde \Pi)\,.
 \ea
This is the decoupling limit of massive gravity around flat space-time and it exhibits all the interactions that arise at or below the scale $\Lambda$. In particular this decoupling limit shows the absence of any BD ghost (the BD ghost would have exhibited itself by interactions of the form
$[\Pi^n]$ which would have led to more than two derivatives at the level of the equations of motion for $\pi$ and would have signalized the presence of an additional excitation, see Refs.~\cite{ArkaniHamed:2002sp,Creminelli:2005qk,Deffayet:2005ys}.)
We now move onto the description of massive gravity around another reference metric before using the previous formalism to infer the decoupling limit around de Sitter.

\subsection{BiGravity}

Another major development in massive gravity was then recently put forward by F.~Hassan and R.~Rosen, realizing that the theory of massive gravity proposed in \cite{deRham:2010kj} and summarized above, could be generalized to an arbitrary metric, leading to bi-gravity \cite{Hassan:2011zd}
\ba
\label{BiGravity_Lagrangian}
\mathcal{L}_{\rm biGravity}=\frac{\mpl^2}{2}\sqrt{-g}\left(R[g]-\frac{m^2}{4}\mathcal{U}(g,f)\right)+\frac{M_f^2}{2}\sqrt{-f}R[f]\,,
\ea
where the potential takes the same form as in \eqref{pot} and (\ref{U2}-\ref{U4}) with now $\mathcal{K}^{\mu}_{\nu}(g)=\delta^{\mu}_{\nu}-\sqrt{g^{\mu\alpha}f_{\alpha\nu}}$ instead of $\sqrt{g^{\mu\alpha}\eta_{\alpha\nu}}$, where the metric $f\mn$ is now dynamical.

This theory is free from the BD ghost \cite{Hassan:2011zd} when different types of matter are then allowed to couple to either metric $S_{m,g}=\int \d^4x \sqrt{-g}\L_m(g\mn,\psi_g)$ and $S_{m,f}=\int \d^4x \sqrt{-f}\L_m(f\mn,\psi_f)$, where $\psi_{g,f}$ symbolize species living on the metric $g\mn$ and $f\mn$ respectively. Matter coupling directly to both metric $\L_m(g,f,\psi)$ leads to the re-emergence of the BD ghost.

In this biGravity theory, the Planck scales $\mpl$ and $M_f$  for both metrics are independent, and we recover massive gravity by simply sending one of this Planck scales (say $M_f$) to infinity, so as to decouple the dynamics from one of the metrics (here $f\mn$). Writing the metric $f\mn=\eta\mn+ \chi\mn/M_f$,
in the limit $M_f \to \infty$  (while keeping the scales $\mpl$ and $m$ fixed in that limit),  we recover massive gravity plus a completely decoupled massless spin-2 field $\chi\mn$,
\ba
\mathcal{L}_{\rm biGravity}\xrightarrow{M_f\to \infty}
\L_{MG}+\frac 12 \chi^{\mu\nu}\Ein^{\alpha\beta}\mn \chi_{\alpha\beta}\,,
\ea
where the Lagrangian for massive gravity given in \eqref{MG_lagrangian} remains fully non-linear in this limit and is expressed in terms of the full metric $g\mn$ and the background metric $\eta\mn$.

\subsection{Massive Gravity around de Sitter}

\label{sec:MG_dS}

Let us now consider the specific case where a cosmological constant couples to the metric $f\mn$. To be more precise, let us implement the biGravity action \eqref{BiGravity_Lagrangian} with a  matter sector of the form
\ba
\label{S_matter}
S_{\rm matter}= - M_f^2 \int \d^4x \sqrt{-f} \Lambda_f \,.
\ea
There can also be in principle another cosmological constant living on top of the metric $g\mn$ but for now we only consider $\L_m(g,\chi_g)$ to source perturbations for $g\mn$. The background field equations of motion are then given by
\ba
M_f^2 G\mn[f] + \frac{m^2 \mpl^2}{4\sqrt{-g}} \(\frac{\delta}{\delta f^{\mu\nu}}\sqrt{-g}\, \mathcal{U}(g,f)\)&=&-M_f^2\Lambda_f f\mn\\
\mpl^2 G\mn[g] + \frac{m^2 \mpl^2}{4\sqrt{-g}} \(\frac{\delta}{\delta g^{\mu\nu}}\sqrt{-g}\, \mathcal{U}(g,f)\)&=&0\,.
\ea
Taking now the limit $M_f\to \infty$ while keeping the cosmological constant $\Lambda_f$ fixed, the background solution for the metric $f\mn$ is nothing else but dS (or AdS depending on the sign of $\Lambda_f$). So expressing the metric $f\mn$ as  $f\mn=\gamma\mn+ \chi\mn/M_f$, where $\gamma\mn$ is the dS metric with Hubble parameter $H=\sqrt{\Lambda_f/3}$ and taking the limit $M_f \to \infty$,   we recover massive gravity on (A)dS plus a completely decoupled massless spin-2 field $\chi\mn$,
\ba
\mathcal{L}_{\rm biGravity}- M_f^2 \int \d^4x \sqrt{-f} \Lambda_f\  \xrightarrow{M_f\to \infty}
\
\frac{\mpl^2}{2}\sqrt{-g}\left(R-\frac{m^2}{4}\mathcal{U}(g,\gamma)\right)
+\frac 12 \chi^{\mu\nu}\Ein^{\alpha\beta}\mn \chi_{\alpha\beta}\,,
\ea
where once again the scales $\mpl$ and $m$ are kept fixed in the limit $M_f\to \infty$.
 $\gamma\mn$ now plays the role of a reference metric on top of which excitations for the massive graviton live \cite{Hassan:2011tf}. Here again
the Lagrangian for massive gravity is given in \eqref{MG_lagrangian} with now $\mathcal{K}^{\mu}_{\nu}(g)=\delta^{\mu}_{\nu}-\sqrt{g^{\mu\alpha}\gamma_{\alpha\nu}}$  and remains fully non-linear in this limit $M_f\to \infty$ and is expressed solely in terms of the full metric $g\mn$ and the reference metric $\gamma\mn$. In what follows we will study this theory further by looking at the decoupling limit $\mpl\to \infty,\, m\to 0$ while keeping the scale $\Lambda=(\mpl m^2)^{1/3}$ fixed.

\section{Helicity-0 mode on de Sitter space}
\label{sec:5d}

As seen in the previous section and in particular in paragraph \ref{sec:helicity0_Minkowski}, identifying the helicity-0 mode relies on identifying the space-time with the internal symmetry. Only around a maximally symmetric space-time does it make sense to perform a helicity decomposition of a spin-2 field. Around an arbitrary reference metric, one no longer has a full Poincar\'e or equivalent group, and there is therefore no Poincar\'e representation to talk about. Since (A)dS space-time is also a maximally symmetric manifold, the notion of a massive spin-2 field around this space-time is meaningful, but one requires additional work to fully identify the helicity-0 mode. Around Minkowski, one can easily identify the internal and space-time global symmetry but this is no longer the case around (A)dS. Instead, we will use a similar trick to what was used in \cite{Burrage:2011bt} and embed $d$-dimensional (A)dS into $d+1$-dimensional Minkowski space-time. In $d+1$ dimensional Minkowski, the identification of the different helicity modes is straightforward, and the only subtlety lies in their projection back into $d$-dimensional (A)dS. We will do this explicitly in what follows and focus for definiteness on dS although all the results are easily generalizable to AdS, by identifying $H_{\rm dS}^2= - H^2_{\rm AdS}$.

In this section Greek indices $\mu,\nu,\cdots=0,\cdots,d-1$ represent $d$-dimensional space-time indices, middle of the alphabet roman indices $i,j,k,\cdots=1,\cdots,d-1$ represent space indices, beginning of the alphabet small roman indices $a,b, \cdots=0,\cdots,d-1$ represent $d$ internal indices, whilst capital roman indices $A,B, \cdots=0, \cdots, d$ represent $d+1$ space-time indices and middle of the alphabet capital roman indices $M,N, \cdots= 0, \cdots d$ represent $d+1$ internal indices.

\subsection{Strategy}

As seen in section \ref{sec:MG_dS}, to build the theory of massive gravity around a dS metric, we need to compute the quantity
\ba
M\mupn=g^{\mu\alpha}\g_{\alpha\nu}\,,
\ea
where $g\mn$ is the dynamical metric and $\g\mn$ stands for the dS reference metric. Now since $\g\mn$ is a reference metric, it is fixed and does not transform under coordinate transformations. To re-establish covariance, we wish to use the standard \stu trick and express $\g\mn$ in terms of $d$ \stu fields (as we did for the flat Minkowski reference metric, $\eta\mn \to \tilde \eta\mn=\eta_{ab}\p_\mu \phi^a \p_\nu \phi^b$, where $\tilde \eta\mn$ is now a tensor). However this na\"ive generalization for de Sitter, $\g\mn \to \tilde \g\mn=\gamma_{ab}\p_\mu \phi^a \p_\nu \phi^b$, would not lead to the correct identification of the physical states in the decoupling limit for the reasons indicated in paragraph \ref{sec:helicity0_Minkowski} and at the beginning of this section. When including curvature corrections the helicity-0 mode is no longer a scalar and nor is the field $\tilde \pi$ in the split $\phi^a= x^a  -\p^a \tilde \pi$. Thus there is no longer any reason why this $\tilde \pi$ should correctly identify the helicity-0 mode, and in general it does not, see Ref.~\cite{deRham:2011rn} for further details.

In appendix \ref{appendix:naive} we show explicitly what happens if one uses this na\"ive method to introduce the different degrees of freedom and we show that one is then lead to introduce non-trivial non-covariant field redefinitions at each order in the field expansion.
Instead of using this na\"ive approach and guessing the correct non-covariant diagonalization between the helicity-2 and -0 fields at each order in the expansion, which would be correct but very pedestrian, we use in what follows some knowledge from the decoupling limit in Minkowski. This is made possible by embedding the $d$-dimensional de Sitter spacetime into $d+1$-dimensional Minkowski, \ie we write $\g\mn\d x^\mu \d x^\nu= \(\eta_{AB}\d Z^A \d Z^B\)|_{\text{projected on d-dim dS hypersurface}}$ (this is similar in spirit to the method used in \cite{Joung:2012rv}).
It is then easy to ``covariantize" the $d+1$-dimensional reference metric: $\eta_{AB}\to \tilde \eta_{AB}= \eta_{MN}\p_A \phi^M \p_B \phi^N$, in terms of $d+1$ \stu fields $\phi^M$. Once this covariantization is done, we then ``simply" identify its projected $d$-dimensional counterpart:
\ba
\label{Covariant_Metric}
\tilde \g\mn\d x^\mu \d x^\nu&=& \(\tilde \eta_{AB}\d Z^A \d Z^B\)|_{\rm projected}=\(\eta_{MN}\p_A \phi^M \p_B \phi^N\d Z^A \d Z^B \)|_{\rm projected}\nn \\
&=&\eta_{MN}\p_\mu \phi^M \p_\nu\phi^N \d x^\mu \d x^\nu.
\ea
As we will see below, we can then identify the scalar degree of freedom $\P$ by setting $\phi^M=Z^M-\eta^{MN}\p_N \P$. This will then give us the expression for $\tilde \g\mn$ in terms of the helicity-0 mode $\P$ as identified in the decoupling limit.

\subsection{Embedding of $d$-dimensional de Sitter into $d+1$-dimensional Minkowski}

We start with the $d$-dimensional dS manifold $\cal C$,
\be
\gamma_{\mu \nu} \d x^{\mu} \d x^{\nu}=-\d t^2+ e^{2 H t} \d{\boldsymbol{x}}^2\,,
\ee
embed it into $d+1$-dimensional Minkowski space in de Sitter slicing with coordinates $\lbrace X^{A} \rbrace=\lbrace X^{\mu}=x^{\mu},X^d=Y \rbrace$ and denote by $G_{AB}$ the five-dimensional Minkowski metric in this slicing:
\be
\d s^2=G_{AB}\d X^A\d X^B=e^{-2 H Y} \left( \d Y^2+ \g_{\mu \nu} \d x^{\mu} \d x^{\nu}\right)\,,
\ee
so that we recover $d$-dimensional dS space at the hypersurface $Y=0$.
Equivalently, we may work in the standard flat-slicing of Minkowski (with corresponding metric $\eta_{AB}$),
\be
\d s^2=\eta_{AB} \d Z^A \d Z^B\,,
\ee
where the Minkowski coordinates $\lbrace Z^{A} \rbrace$, are expressed in terms of the ``de Sitter" slicing ones as
\bea
Z^0&=&e^{-H Y} \left( H^{-1} {\rm sh}(Ht) +\half e^{H t} H  {\boldsymbol{x}}^2  \right) \label{Z0} \\
Z^d&=& e^{-H Y} \left( H^{-1} {\rm ch}(Ht) -\half e^{H t} H  {\boldsymbol{x}}^2  \right) \label{Z4} \\
Z^i&=&e^{- H Y} e^{Ht} x^i\,, \quad i=1,2,d-1 \,. \label{Zi}
\eea
In the coordinate system $\{Z^A\}$,  $\cal C$ is identified as the hypersurface of constant radius $H^{-1}$
\be
\eta_{AB} Z^A Z^B=H^{-2} \,.
\label{constraint1}
\ee

\subsection{Helicity-0 mode}

We now introduce $d+1$ \stu fields  $\phi^M$, living in $d+1$-dimensional Minkowski space, which we split as
\be
\phi^M=Z^M- \tilde V^M\,.
\ee
When projected onto the $Y=0$ or equivalent \refeq{constraint1} hypersurface, the $d+1$ \stu fields also satisfy the relation,
\ba
\eta_{MN} \phi^M \phi^N=H^{-2}\,,
\ea
which implies for $\tilde V^M$
\ba
\label{const}
-2 Z_M \tilde V^M+\eta_{MN} \tilde V^M\tilde V^N=0\,.
\ea
To avoid any confusion, the vector $\tilde V^A$ introduced so far in the flat slicing of $d+1$-dimensional Minkowski is denoted by $V^A$ in the de Sitter slicing of Minkowski, more precisely the relation between $\tilde V^A$ and $V^A$ satisfies the standard vector change of coordinates:
\ba
\eta_{AB}\tilde V^A \d Z^B = G_{AB}V^B \d X^B\,,
\ea
and identifying internal with space-time indices, we have
\ba
Z_A \tilde V^A = -\frac{e^{-2 HY}}{H} V^{Y} \hspace{20pt}{\rm and}\hspace{20pt} \eta_{MN}\tilde V^M\tilde V^N= \g_{MN}V^M V^N\,,
\ea
and the relation \eqref{const} therefore simplifies to
\be
\label{V^4}
\frac{2}{H} V^{Y} + \left( (V^{Y})^2+V^2 \right)=0\,,
\ee
with $V^2=\g_{\mu \nu} V^\mu V^\nu$.
The relevant solution for $V^Y$ is then
\be
V^{Y}=-\frac{1}{H} \(1- \sqrt{1 - H^2 V^2}\)\,.
\label{explicit-Piy}
\ee
As explained previously in Eq.~\eqref{Covariant_Metric}, the $d$-dimensional covariantized reference metric is expressed in terms of the \stu fields as
\bea
\tilde \g\mn &=&\eta_{AB}\p_\mu \phi^A\p_\nu \phi^B \\
&=& \g_{\mu \nu}-2 \partial_{(\mu} Z_A \partial_{\nu)}\tilde V^A+\eta_{AB} \partial_{\mu}\tilde V^A \partial_{\nu}\tilde V^B\,.
\label{def-f}
\eea
Expressing \eqref{def-f} in terms of the coordinates $\{X^A\}$, and using the relation \eqref{explicit-Piy}, we find (see appendix \refeq{sec:Coord-transformation})
\be
\tilde \g\mn =\g_{\mu \nu}-S_{\mu \nu}-S_{\nu \mu}+ S_{\mu \alpha} \g^{\alpha \beta} S_{\nu \beta} +\frac{H^2}{1- H^2 V^2} \T_{\mu} \T_{\nu}\,,
\label{explicit-f}
\ee
with
\be
S_{\mu \nu}=\nabla_{\mu} V_{\nu}+\g_{\mu \nu}  \left(1-\sqrt{1-H^2 V^2 } \right)
\ee
and
\ba
\T_{\mu}= \frac 12 \p_\mu V^2 -  \sqrt{1-H^2 V^2 }\,  V_{\mu}\,
\label{def-Y}
\ea
and where all the covariant derivatives are with respect to $\g_{\mu \nu}$. The expression for the ``tensor fluctuation" $H\mn=g\mn-\tilde \g\mn$  is then
\be
H_{\mu \nu}=g_{\mu \nu}-\g\mn+S_{\mu \nu}+S_{\nu \mu}- S_{\mu \alpha} \g^{\alpha \beta} S_{\nu \beta} -\frac{H^2}{1-H^2 V^2} \T_{\mu} \T_{\nu}\,.
\label{def-H}
\ee
At this stage, we may split $V^\mu$ into
\ba
V_\mu = A_\mu +\p_\mu \P\,,
\ea
where $A_\mu$ now describes the helicity-1 mode and is a vector field in the decoupling limit while $\P$ is a scalar field in the decoupling limit and successfully encodes  the helicity-0 mode. In the decoupling limit we will consider, the helicity-1 mode always arises quadratically (see refs.~\cite{deRham:2010gu} and \cite{deRham:2010ik}), and can hence be consistently set to zero. We emphasize that beyond the decoupling limit this result is no longer valid and the helicity-1 mode should be fully investigated. Furthermore when exploring the decoupling limit around Anti de Sitter, the helicity-1 mode plays a crucial role and remains very alive in the decoupling limit. This aspect will be soon presented elsewhere \cite{AdS}. To simplify the derivation, we dedicate the bulk of the paper to the helicity-0 mode and therefore set the helicity-1 mode to zero.

Finally, we point out that in the Minkowski limit where $H\to 0$ keeping all other scales fixed, we recover the standard expression of the covariantized metric in terms of the helicity-0 mode $\P$,
\ba
\tilde \eta\mn =\eta_{\mu \nu}-2 \p_{\mu} \p_{\nu} \P+  \eta^{\alpha \beta} \p_{\mu} \p_{\alpha} \P\p_{\nu} \p_{\beta} \P\,.
\label{CovMetricMinkoski}
\ea

\section{Fierz-Pauli on de Sitter and Higuchi Bound}
\label{sec:Linearized}

In this section, we focus on the linearized part of the Massive Gravity action \eqref{MG_lagrangian}, which is equivalent to the Fierz-Pauli action when expanded around de Sitter, $g\mn = \g\mn + \tilde h\mn = \g\mn + h\mn / \mpl^{(d-2)/2}$. At the linearized level only the potential $\L_{\rm der}^{(2)}$ contributes in \eqref{MG_lagrangian}
\ba
\L^{(2)}_{\rm MG, \, dS}=
-\frac {1}4 h^{\mu\nu}(\Ein_{\rm dS})^{\alpha\beta}\mn\, h_{\alpha\beta} - \frac {\mpl^{d-2}m^2}8  \g^{\mu\nu}\g^{\alpha \beta}\(H_{\mu\alpha}H_{\nu\beta}
-H\mn H_{\alpha \beta}\)\,,
\ea
where $\Ein_{\rm dS}$ is now the Lichnerowicz operator on de Sitter,
\ba
(\Ein_{\rm dS})^{\alpha\beta}\mn\, h_{\alpha\beta}
&=&-\frac 12 \Big[\Box  h\mn-2\nabla_{(\mu} \nabla_\alpha h^\alpha_{\nu)}+\nabla_\mu\nabla_\nu  h-\g \mn (\Box h-\nabla_\alpha\nabla_\beta  h^{\alpha\beta})\\
&& \hspace{20pt}+2(d-1)H^2 \(h\mn- \frac 12 h\g\mn\)\Big]\,.\nn
\ea
Using the expression \eqref{def-H}, which at the linearized level simplifies to $H\mn=\tilde h\mn+2 \nabla_\mu \nabla_\nu \P$, we obtain
\ba
\label{L_2_ourway}
\L^{(2)}_{\rm MG, \, dS}&=&
-\frac {1}4 h^{\mu\nu}(\Ein_{\rm dS})^{\alpha\beta}\mn\, h_{\alpha\beta} - \frac{m^2}8
\(h\mn^2-h^2\)\\
&&-\frac {m^2}2 h^{\mu\nu}\(\bar \Pi\mn - [\bar \Pi] \g\mn\)-\frac{m^2}{2}\([\bar \Pi^2]-[\bar \Pi]^2\)\,,\nn
\ea
where we use here the notation $\bar \Pi\mn=\nabla_\mu \nabla_\nu \bar \pi$, and have normalized the field $\bar \pi=\mpl^{(d-2)/2} \P$.
After integration by parts,
$[\bar \Pi^2]=[\bar \Pi]^2-(d-1) H^2 (\p \bar \pi)^2$.

The helicity-2 and -0 modes are diagonalized by setting
$h\mn= \bar h\mn + \frac{2  m^2}{d-2} \bar \pi \g\mn$:
\ba
\label{quadratic}
\L^{(2)}_{\rm MG, \, dS}&=&
-\frac {1}4\bar h^{\mu\nu}(\Ein_{\rm dS})^{\alpha\beta}\mn\, \bar h_{\alpha\beta}-\frac{m^2}{8}\(\bar h\mn^2-\bar h^2\)\\
&&-\frac{m^2}{2}\frac{d-1}{d-2}\(m^2-(d-2)H^2\) \(\(\p \bar \pi\)^2 - \bar h \bar \pi -\frac{d}{d-2}\bar \pi^2\) \,.\nn
\ea
As to the coupling to matter, the loss of general covariance (or four-dimensional diffeomorphism invariance at the linear level) implies, at least in principle, that  matter can couple differently to the metric, and external matter is not necessarily conserved (the conservation of energy is only a requirement in the exact massless case where the symmetries of GR impose it). With this in mind, $\pi$ can therefore couple not only to the trace of the stress-energy tensor of external matter sources, but as well to other combinations of it. For instance, one way matter could couple in this theory is through
\ba
\label{Lmat}
\L_{\rm mat}= \frac 1{2} H\mn T^{\mu\nu}=  \frac 1{\mpl^{(d-2)/2}} \(\frac 12 \bar h\mn T^{\mu\nu}+\frac{m^2}{(d-2)} \bar \pi T+ \bar \pi \, \nabla_\mu \nabla_\nu T^{\mu\nu}\)
\ea
where the last term may not necessarily vanish. More generally, we could consider that the helicity-0 mode couples to $\mathcal{T}$, where
$\mathcal{T}$ could include for instance $T$, $\nabla_\mu \nabla_\nu T^{\mu\nu}$ {\it etc}.

We now see the standard results of massive gravity on de Sitter:
\begin{enumerate}

\item In the massless limit $m\to 0$, the helicity-0 mode decouples from conserved matter, as is clear from \eqref{Lmat}, which suggests the absence of the standard  vDVZ (van Dam-Veltman-Zakharov) discontinuity, \cite{Kogan:2000uy,Porrati:2000cp}. To be more precise, if we assume $\nabla_\nu T^{\mu\nu}=0$, then after the field has been canonically normalized,
    \ba
    \phi= m\sqrt{m^2-(d-2)H^2}\ \bar \pi\,,
    \ea
    the coupling to matter scales as
            \ba
    \mathcal{L}_{\rm mat}^{(\phi)}=\frac{1}{d-2} \frac{m}{\sqrt{m^2-(d-2)H^2}}\,  \frac{1}{\mpl^{(d-2)/2}}\,\phi \mathcal{T}\,,
    \ea
    so this coupling tends to zero in the limit $m\to 0$ provided the Hubble parameter $H$ (or equivalently the cosmological constant $\Lambda_f$) vanishes more slowly in that limit (\ie $m/H\to 0$).

    This result was also proven in the Newtonian approximation in \cite{Deser:2001gz}.
    (The absence of discontinuity around a cosmological background was also pointed out for spin-3/2 fields, \cite{Deser:2000de}).
\item As is clear from the helicity-0 kinetic term, the theory only makes sense for $m^2=0$ (\ie GR) or $m^2>(d-2)H^2$, which corresponds to the well-know Higuchi bound, \cite{Higuchi:1986py,Deser:2001wx} (see \cite{Gabadadze:2008ha} for an attempt to eliminate this ghost at the linearized level through kinetic mixing in a slightly different model). Another possibility is $m^2<0$, in which case the field is not only tachyonic, but the helicity-1 field is then a ghost. (At quadratic order the helicity-1 mode comes in as $-\frac{m^2}{4}F\mn^2$).

    The presence of this Higuchi bound makes the previous resolution of the vDVZ discontinuity obsolete. Indeed, in the massless limit $m\to 0$, one should keep $m^2>2 H^2$, and so the limit should be taken with $H\to 0 $ simultaneously (at least as fast as $m$).

    This is of course not the case in Anti de Sitter \cite{Porrati:2000cp}, where there is no such Higuchi bound, but the absence of vDVZ discontinuity does not prevent the fact that interactions are important at a low energy-scale and ought to be resumed \cite{Deffayet:2001uk}, as we shall show elsewhere \cite{AdS}.

\item The special case $m^2=(d-2)H^2$ ought to be treated with care as the helicity-0 mode then loses its kinetic and mass term. This corresponds to the partially massless (PM) case as proposed by S.~Deser and A.~Waldron, \cite{Deser:2001wx,Deser:2001us,Deser:2001xr} where the helicity-0 excitation was argued to disappear. This special case is all the more interesting as it is conformally invariant, \cite{Deser:2004ji} and could have an interesting phenomenology, \cite{Deser:2006zx}. We will pay particular attention to this special case in the paper and show that there exists a unique non-linear completion of PM gravity for which the helicity-0 completely disappears, at least within the decoupling limit, in the absence of matter. The helicity-0 mode only manifests then via its coupling to matter and imposes the constraint
    \ba
    \label{Matter-constraint}
    \nabla_\mu \nabla_\nu T^{\mu\nu}=-\frac{m^2}{d-2}T\,,
    \ea
    so that the failure of energy conservation is proportional to the graviton mass.

\end{enumerate}

\section{Decoupling Limit on de Sitter}
\label{sec:DL}

\subsection{Interaction scales}

As explained in the previous section, the existence of the Higuchi bound in dS makes it impossible  to consider the limit $m\to 0$ without simultaneously sending $H\to 0$ at least at the same rate.
One could of course also send $H/m\to 0$ in that limit, but would then recover the decoupling limit of massive gravity around flat space \cite{deRham:2010ik,deRham:2010kj}, and would then loose any information about the PM special case for which $H^2$ is related to $m^2$.
So to derive the decoupling limit, we send $\mpl \to \infty$, $m\to 0$, while keeping both the scale $\Lambda=(m^2\mpl^{(d-2)/2})^{2/(d+2)}$ and the ratio $H/m$ fixed in that limit.  In this limit we keep the canonically normalized field finite:
\ba
h\mn = \mpl^{(d-2)/2} \tilde h\mn \to {\rm const}\hspace{10pt}{\rm and}\hspace{10pt}
\pi = \mpl^{(d-2)/2} m^2  \P =\Lambda^{(d+2)/2} \P \to {\rm const}\,.
\ea

Focusing on the massive term ${\cal U}$, prior to performing any integrations by parts, $h$  always appear with no derivatives and $\P$ appear in two forms, either with two derivatives, $\partial^2 \P$, or in the combination $H^2 (\partial \P)^2$, so the mass term involves arbitrary powers of $\tilde h$, $(\p^2 \P)$ and $H^2 (\p \P)^2$, which we write schematically as
\ba
\mpl^{d-2} m^2 \sqrt{-g}\, \mathcal{U}&\sim& \mpl^{d-2} m^2 \sum_{k,{n}, \ell \ge0}c_{k,\ell,n }\left(\frac{h}{\mpl^{(d-2)/2}} \right)^{k} \(\frac{\p^2 \pi}{\Lambda^{(d+2)/2}}\)^{{n}} \(\frac{H^2 (\p \pi)^2}{\Lambda^{d+2}}\)^{\ell }\\
&\sim& \sum_{k,{n}, \ell \ge0}c_{k,\ell,n }\, (\mpl^{\frac{d-2}{2}(1-k)}H^{2\ell })\ \frac{h^k (\p^2 \pi)^{n} (\p \pi)^{2\ell }}{\Lambda^{\frac{d+2}{2}(2\ell + {n}-1)}}\,.
\label{interactions}
\ea
All terms with $\ell > 1-k$ drop out in the decoupling limit, and so the only three non-trivial class of terms are the one for which
\begin{itemize}
\item $k=\ell =0$, \ie the ones schematically of the form $\mpl^{(d-2)/2} \(\p^2 \pi \)^{n}/\Lambda^{(d+2)(n-1)/2}$. Na\"ively these terms diverge in the decoupling limit, meaning that one should work at a lower energy scale. However the special Ghost-free theory of massive gravity presented in section \ref{sec:MG} \cite{deRham:2010ik,deRham:2010kj}, has been carefully engineered so that such terms only appear as total derivatives up to integrations by parts. Whilst these integrations by parts are irrelevant on Minkowski, on de Sitter on the other hand they lead to some curvature contributions which cannot be ignored. More precisely, these terms can be integrated by parts as
      \ba
    \label{L00}
     \L_{\rm der}^{({n})}(\tilde \Pi) &\equiv &H^2 \L^{({n})}_{(k=0,\ell =0)} = -\frac{1}{(d-{n})!} \,  \mathcal{E}^{\alpha_1 \cdots \alpha_d}
    \mathcal{E}^{\beta_1 \cdots\beta_n }{}_{\alpha_{n +1} \cdots \alpha_d}\ \tilde \Pi_{\alpha_1 \beta_1}\cdots \tilde \Pi_{\alpha_{n}\beta_{n}}\\
&=& \frac{{n}-1}{(d-{n} )!} \, \mathcal{E}^{\alpha_1 \cdots \alpha_d}\mathcal{E}^{\beta_1 \cdots\beta_n }{}_{\alpha_{n +1} \cdots \alpha_d}\ \p_{\alpha_1}\P \tilde \Pi_{\alpha_2 \beta_2}\cdots \nabla_{\beta_1}\nabla_{\alpha_{n}}\nabla_{\beta_{n}}\P
    \ea
   (where we recall that $n\le d$).
    Rewriting $\nabla_{\beta_1}\nabla_{\alpha_{n}}\nabla_{\beta_{n}}\P $  as $\nabla_{\alpha_{n}}\nabla_{\beta_1}\nabla_{\beta_{n}}\P + R^\mu_{\  \beta_{n} \beta_1 \alpha_{n} }\p_\mu \P$, we can see that the first term cancels after contraction with the antisymmetric Levi-Cevita tensor. The Riemann curvature is to be evaluated on the de Sitter reference metric and gives rise to
    \ba
        \L^{({n})}_{(k=0,\ell =0)}  &=& ({n}-1)(d-{n}+1) \, \mathcal{E}^{\alpha_1 \cdots \alpha_d}\mathcal{E}^{\beta_1 \cdots\beta_{n-1} }{}_{\alpha_{n}\cdots \alpha_d}\ \p_{\alpha_1}\P \p_{\beta_{1}} \P \, \tilde \Pi_{\alpha_2 \beta_2}\cdots \tilde \Pi_{\alpha_{{n}-1} \beta_{{n}-1}}\nn \\
&=& \frac{{n}}{2}({n}-1)(d-{n}+1)  \, (\p \P)^2 \L^{({n}-2)}_{\rm der}(\tilde \Pi) \propto \L^{({n})}_{\rm Gal}\,. \label{L der}
    \ea
     We therefore obtain a Galileon interaction, \cite{Deffayet:2009mn}, which arises at the finite scale \\ $\mpl^{(d-2)/2} H^2 / \Lambda^{(d+2)(n-1)/2} \sim 1/\Lambda^{(d+2)(n-2)/2}$, is hence relevant in the decoupling limit and contributes as
      $\sqrt{-g}  \L^{({n})}_{\rm der}(\K) \supset H^2 {\cal L}_{(k=0,\ell =0)}^{({n})}$. It is worth pointing out that it is precisely due to that contribution that the kinetic term for $\pi$ acquires a new piece proportional to $(d-1)H^2$ in \eqref{quadratic}.
\item The second class of interactions which are relevant in the decoupling limit are the ones for which $k=1$ and $\ell=0$, \ie the ones schematically of the form $\tilde h (\partial^2 \P)^{n}$, which are the ones already accounted for in the decoupling limit of massive gravity around flat space, \cite{deRham:2010ik,deRham:2010kj}. They  take precisely the same form here\footnote{Notice that since we take the limit $H\to 0$, in the decoupling limit, the fields still live around Minkowski.}:
    \ba
    {\cal L}_{(k=1,\ell =0)}^{({n})}& \equiv &  h\mn \frac{\delta}{\delta h\mn}\(\sqrt{-g}\L^{({n})}_{\rm der}(\K)\)_{\lvert h\mn=0, H^2=0}\\
    & = &  h^{\mu\nu} \(X\mn^{({n})}+{n} X\mn^{({n}-1)}\)\,,\label{L10}
    \ea
where $\sqrt{-g}  \L^{({n})}_{\rm der}(\K) \supset {\cal L}_{(k=1,\ell =0)}^{({n})}/\mpl^{(d-2)/2}$, and $X\mn^{(n)}$ is given in expression \eqref{Xmn}. Note also that $X\mn^{(n)}$ is identically $0$ for $n \geq d$ in $d$ dimensions.
\item Last but not least, the most important class of interactions which depends crucially on correctly identifying the helicity-0 mode around de Sitter is the one for which  $k=0$ and $\ell =1$,  \ie the interactions of the form $H^2 (\partial \P )^2 (\partial^2 \P)^{n}$, which also take a Galileon form. These terms are derived explicitly in appendix \ref{sec:appendix dS}:
    \ba
\mathcal{L}^{({n})}_{(k=0,\ell =1)}&\equiv& \frac{\delta}{\delta H^2} \left(\sqrt{-g}\, \L_{\rm der}^{({n})}(\K) \right)_{\lvert h\mn=0, H^2=0}\\
&=&- \sum_{m=1}^{{n}}f^{{n},m}(\P)\  \L_{\rm der}^{({n}-m)} (\tilde \Pi)\,,
\label{L01}
\ea
with $\L_{\rm der}^{(0)}=1$, $[\phi^k] \equiv \partial \P \cdot \tilde \Pi^k \cdot \partial \P$ and
\ba
\label{coeff2}
f^{n,m}(\P)=\frac{(-1)^m}{2} \frac{n!}{(n-m)!} \([\phi^m]-[\phi^{m-1}]+(\partial \P)^2[\tilde \Pi^{m-1}]\)\,.
\ea
They contribute as  $\sqrt{-g}  \L^{({n})}_{\rm der}(\K) \supset H^2 {\cal L}_{(k=0,\ell =1)}^{({n})}$.
\end{itemize}
It is worth emphasizing that all of these results are valid in an arbitrary number of dimensions. Even though the contributions from the last type of terms $(k=0,\, \ell=1)$ do not have any explicit dependance on the dimensionality, it will appear once we make use of the relation $[\tilde \Pi^0]=d$ in \eqref{coeff2}.

Applying the relation \eqref{coeff2}, we find after integrations by parts (we recall that in this limit, the field $\pi$ lives on flat space and derivatives can therefore be commuted at will),
\bea
\mathcal{L}^{({2})}_{(k=0,\ell =1)}= \(d-\frac52\) \L_{\rm Gal}^{(3)} - \frac12 \L_{\rm Gal}^{(4)}  \,.
\eea
Following the same logic for the following terms, we get
\ba
\label{L dS}
\mathcal{L}^{({n})}_{(k=0,\ell =1)}=  \frac{n}{2}\(d-\frac{3n-1}{2}\)  \L_{\rm Gal}^{(n+1)}
- \frac{n}{4}\L_{\rm Gal}^{(n+2)} \,.
\ea
In $d$ dimensions, only the Galileon Lagrangians till $n=d+1$ contribute, $\L_{\rm Gal}^{(d+2)}$ is a total derivative and all the other higher Galileons vanish identically.
\subsection{Full decoupling limit}

Using the previous results, the mass term is thus given in the decoupling limit by
\ba
{\cal L}_{{\rm mass}}& = & \sum_{n=2}^d \frac{\alpha_n}{2} \(\Lambda^{(d+2)} \frac{H^2}{m^2}\(\L^{(n)}_{(k=0, \ell=0)}+ \L^{(n)}_{(k=0, \ell=1)}\)
+\Lambda^{(d+2)/2} \L^{(n)}_{(k=1, \ell=0)}\)\,,
\label{Lmass tot}
\ea
where $\L^{(n)}_{(k, \ell)}$ are given in \eqref{L der}, \eqref{L10} and \eqref{L dS}, and at this stage, all metric contractions can be taken with respect to Minkowski.

Collecting all the results, we finally obtain the full Lagrangian for massive gravity around de Sitter, including all the interactions between the helicity-2 and -0 modes that arise at or below the energy scale $\Lambda$, and for any dimension $d$:
\ba
\L_{{\rm DL}}^{d}&=& -\frac14 h^{\mu \nu} \Ein^{\alpha \beta}\mn h_{\alpha \beta}
+\frac12 h^{\mu \nu} \sum_{n=1}^{d-1}(\alpha_n+(n+1) \alpha_{n+1}) \frac{X^{(n)}\mn }{ \Lambda^{(d+2)(n-1)/2}} \\
&& +\sum_{n=2}^{d+1} \beta \lambda_n  \frac{\L^{(n)}_{\rm Gal}  }{ \Lambda^{(d+2)(n-2)/2} }\,,\nn
\ea
where $X^{(n)}\mn$ and $\L^{(n)}_{\rm Gal}$ are expressed in terms of the canonical field $\pi$, $\lambda_n$ can be usefully written as
\ba
\lambda_n= \frac14 (d-n+1)(n-1)(\alpha_{n-1}+n \alpha_n)-\frac18(n-2)(\alpha_{n-2}+(n-1) \alpha_{n-1}) \,,
\ea
and $\beta=H^2/m^2$, so that none of the terms on the second line are present in Minkowski (we also defined $\alpha_0=\alpha_1=\alpha_{d+1}=0$).

One can partially diagonalize both fields by use of the transformation
\ba
h\mn=\bar h\mn +\frac{2}{d-2} \pi \eta\mn -\frac{1+3 \alpha_3}{\Lambda^{(d+2)/2}}\p_\mu \pi\p_\nu \pi\,,
\ea
yielding, after a long but straightforward calculation:
\ba
\L_{{\rm DL}}^{d}&=& -\frac14 \bar h^{\mu \nu} \Ein^{\alpha \beta}\mn \bar h_{\alpha \beta}
+\frac12 \bar h^{\mu \nu} \sum_{n=3}^{d-1}(\alpha_n+(n+1) \alpha_{n+1}) \frac{X^{(n)}\mn }{ \Lambda^{(d+2)(n-1)/2}}  \\
&& +\sum_{n=2}^{d+1} c_n  \frac{\L^{(n)}_{\rm Gal}  }{ \Lambda^{(d+2)(n-2)/2} } \,,
\label{final}
\nn
\ea
where
\ba
c_n&=&\beta \lambda_n -\frac12 \frac{d-1}{d-2} \delta_{n,2} -\frac34(1+3 \alpha_3) \delta_{n,3}- \left(\frac14 (1+3 \alpha_3)^2+\frac{d-3}{d-2}(\alpha_3+4 \alpha_4) \right) \delta_{n,4} \nonumber \\
&&-\,\, ({\rm if} \,\, n \geq 5)\,\, \frac{n}{8}\left(  (1+3 \alpha_3)(\alpha_{n-2}+(n-1)\alpha_{n-1})+2 \frac{d-n+1}{d-2}(\alpha_{n-1}+n \alpha_n)   \right)\,.
\ea

In any dimensions we see that we recover a decoupling limit which at first sight resembles almost identically to that in Minkowski. The  only difference arises from the new factors of $\beta=H^2/m^2$ which of course will make a crucial difference for PM gravity. However before exploring PM gravity in details, let us first confirm that the decoupling limit is free of any ghost pathologies as long as we stay within the Higuchi bound, $\beta < 1/(d-2)$. This decoupling limit also allows us to explore the Vainshtein mechanism through which the helicity-0 mode decouples from the rest of matter in the massless limit $m,\Lambda \to 0$. The details of this Vainshtein mechanism depend on the respective signs of the different interactions (as well as on the existence of the non-diagonalizable term $X^{(n\geq 3)}\mn$ in \eqref{final}), \cite{Chkareuli:2011te,Koyama:2011yg,Sjors:2011iv,Sbisa:2012zk,Wyman:2011mp} and the new parameter $\beta$ could therefore slightly change the phenomenology. For instance as soon as $\beta\ne 0$ the minimal model\footnote{The minimal model is the one which bears no interactions in the decoupling limit around Minkowski. It is defined with the coefficients $\alpha_n+(n+1)\alpha_{n+1}=0$ for all $2 \le n \le d-1$, \cite{deRham:2010ik,Hassan:2011vm}. For PM, on the other hand, $\alpha_3$ does not satisfy this property but is instead given by eq.(6.2).} acquires non-trivial interactions in the decoupling limit and its Vainshtein mechanism may then be explored in a very similar way as for any other parameters. However rather than exploring the phenomenology of the phase space for de Sitter, we will focus in what follows onto a very specific case, namely that of PM gravity.

\section{Partially Massless Gravity}
\label{sec:PM}

\subsection{The full non-linear theory}
\label{sec:PM-d}

Within the linearized regime, one can see the helicity-0 mode disappearing for the very specific choice $\beta=  1/(d-2)$. We now see that this result remains valid to all orders in the decoupling limit as long as we take the specific choice of parameters:
\ba
\beta&=&\frac{1}{d-2}
\label{beta} \\
\alpha_3&=&-\frac13 \frac{d-1}{d-2}
\label{alpha_3} \\
\alpha_n&=&-\frac{1}{n} \alpha_{n-1}\, \, {\rm for} \, \, n \geq 4\,.
\label{alpha_n}
\ea
All these parameters enter explicitly in the original action. The parameters $\alpha_n$ can be found in \eqref{MG_lagrangian} or its bi-Gravity counterpart \eqref{BiGravity_Lagrangian} through the potential \eqref{pot}. The parameter $\beta$ on the other hand enters in the action through the matter content \eqref{S_matter} for the auxiliary (frozen) de Sitter metric $f\mn$. $\beta$ is related to the cosmological constant $\Lambda_f$ by the relation $\beta=\frac{2 \Lambda_f}{(d-1)(d-2)m^2}$.

With these choices of parameters (\ref{beta} - \ref{alpha_n}) we have entirely identified the fully non-linear completion of the PM theory of gravity, if it exists. It is worth pointing out that PM gravity is very distinct from the minimal model around Minkowski.  For the minimal model on Minkowski, all the interactions for the helicity-0 mode disappear in the decoupling limit but the kinetic term is still fully present and the helicity-0 mode is therefore a fully propagating degree of freedom in that case. In particular we expect its interactions to reappear when dealing with energies above $\Lambda$. For PM on the other hand, even the kinetic term  mode completely disappears. The helicity-0 mode is therefore altogether absent (at least in the decoupling limit), and the PM theory only propagates $d(d-1)/2-2$ degrees of freedom, consisting in $d(d-3)/2$ helicity-2 modes and $d-2$ helicity-1 modes.

At the linear level, the absence of the helicity-0 mode can be understood by the presence of a new symmetry which is a combination of a special linearized diffeomorphism and a conformal transformation \cite{Deser:2001us,Deser:2001xr}:
\ba
\label{sym}
h\mn \to h\mn + 2 \nabla_\mu \nabla_\nu \zeta (x) + \frac{2 m^2}{d-2} \zeta(x) \g\mn
\ea
(which is precisely the way $\tilde \pi$ enters the definition of $H\mn$ at the linearized order and since $\tilde \pi$ disappears from the action, we may immediately conclude that such a transformation is a symmetry of the the theory). The cubic extension of this symmetry was then provided by Y.~.M.~Zinoviev in \cite{Zinoviev:2006im}. Whilst the fully non-linear realization of this symmetry, if it exists, is yet unknown and requires an analysis well beyond the scope of this manuscript, the mere fact that the helicity-0 mode entirely disappears from the decoupling limit is already very suggestive.
In particular we can already conclude that this symmetry generalizes non-linearly (in the gauge parameter, not in the field) in the decoupling limit to (for convenience, we have rescaled $\zeta$ here compared to \eqref{sym})
\ba
&&\frac{h\mn}{\Lambda^{(d+2)/2}} \to \frac{h\mn}{\Lambda^{(d+2)/2}}+ \frac{1}{m^2} \left( 2 \nabla_\mu \nabla_\nu \zeta -( \nabla_\mu \nabla_\nu \zeta)^2  \right)  \\
&+&\frac{1}{d-2} \left(2 \zeta \gamma\mn + \nabla_\mu \zeta \nabla_\nu \zeta  + (\nabla \zeta)^2 \(\gamma\mn- \nabla_\mu \nabla_\nu \zeta\) - \(\nabla_\mu \nabla_\alpha \zeta
 \nabla^{\alpha} \zeta  - \nabla_\mu \zeta\)\(\nabla_\nu \nabla_\beta \zeta \nabla^{\beta} \zeta  - \nabla_\nu \zeta\) \right) , \nonumber
\ea
which might give a hint onto how the symmetry gets generalized fully non-linearly (once again if it exists)\footnote{We thank Kurt Hinterbichler for useful discussions about this point.}.

\subsection{Symmetry and Counting in the Full Theory}

At this point one may stop and wonder why this symmetry, if it really exists, has not yet been discovered in other studies which analyze the constraint system of the full theory \cite{deRham:2010kj,Hassan:2011hr,Hassan:2011ea,Hassan:2011tf}. The reason for this, is that these previous studied showed the existence of two constraints, but did not show that there was never any additional constraint in the general realization of the theory. Instead, based on the knowledge that the theory had already five propagating degrees of freedom at the linear level one can conclude that any non-linear realization of this theory which is smoothly connected (\ie without infinitively strong coupling issues) to the linear theory necessarily have {\it at least} five propagating degrees of freedom. The presence of two second-class constraints in \cite{deRham:2010kj} not only at the linear level but also beyond is thus sufficient to prove that the theory never excites {\it more} than  these five degrees of freedom.

In the very special case of PM gravity, the theory has already lost one degrees of freedom at the linear level as one see from \eqref{quadratic}, so one cannot rely on the linear theory to infer the absence of an additional constraint beyond the two second-class constraints already found in \cite{Hassan:2011hr,Hassan:2011ea}. Instead one should check explicitly in this case whether the constraint algebra closes or whether an additional constraint is generated. Since this has not been yet been performed explicitly for massive gravity on de Sitter, there is no tension between the potential existence of a symmetry in PM gravity and the constraint algebra found in the full theory.

If a symmetry is present the counting in the phase space Hamiltonian language should go as follows: In unitary gauge, the space components metric $\gamma_{ij}$ and its momentum conjugate lead to 12 degrees of freedom (dofs). The two second-class constraints (not associated to a symmetry) found in \cite{Hassan:2011hr,Hassan:2011ea} then each remove 1 dof. In PM then should then be an additional tertiary constraint which should be first-class and thus remove 2 additional dofs, leading to a total of $12-1-1-2=8=4\times 2$, \ie 4 physical dofs.

In the \stu language, in physical space the counting goes instead as follows: Starting with the 10 dofs in the metric and the 4 dofs in the \stu fields, we have 14 physical dofs. However the \stu fields restore a copy of diffeomorphism invariance  which removes $4\times 2$ dofs. Furthermore in PM gravity the existence of a potential new symmetry removes itself 2 additional physical dofs, leading to a total of $14-4\times 2- 2= 4$ physical dofs, which is again the correct counting. In the \stu language, the additional PM symmetry thus removes both the helicity-0 mode and its BD companion directly. This is easily understandable as the BD ghost is a parasite that rides on top of the standard helicity-0 mode, so PM gravity that has no helicity-0 mode has no BD ghost either.

\subsection{Vainshtein Mechanism or Absence Thereof}

As we depart ever so slightly from the PM parameters defined previously, for instance as soon as $\beta = (1-\varepsilon)/(d-2)$, (with $\varepsilon>0$), we are back to a ``standard" theory of massive gravity, propagating five degrees of freedom in four dimensions. At least at the linearized level, we therefore expect a vDVZ discontinuity in the limit $\varepsilon \to 0 $, \cite{vDVZ}.
As we shall see below, the resolution of this discontinuity lies on whether or not we approach it when the coefficients $\alpha_n$ are that of the PM theory.

\subsubsection{None PM parameters}

Let us start this discussion by assuming that at least one of the parameters $\alpha_n$ differs from that of the PM theory. For definiteness, let us work in four dimensions and assume $\alpha_3+4 \alpha_4=0$ as is the case for PM, but with $\alpha_3\ne -1/2$. For $\e>0$, the theory propagates five healthy degrees of freedom while at $\e \equiv 0$ the helicity-0 mode is infinitely strongly coupled. Unsurprisingly, the resolution of this apparent discontinuity lies in  the existence of a Vainshtein mechanism, which is already manifest in this decoupling limit, \cite{Vainshtein:1972sx}. To best see the effect of the interactions arising in the limit\footnote{We emphasize that this limit is distinct from the PM limit. We are not recovering the PM theory of gravity as $\e \to 0$. Nevertheless, we do recover a theory for which the helicity-0 mode decouples.} $m^2\to 2 H^2$, we canonically normalize the field $\pi$ to $\hat \pi = \sqrt{\varepsilon} \pi$ so that the decoupling limit for that field reads
\ba
\L^{(\pi)}_{\rm DL} =
-\frac 34 (\p \hat \pi)^2-\frac {3}{8 \Lambda^3}\frac{1+2\alpha_3}{\varepsilon^{3/2}} \L^{(3)}_{\rm Gal}(\hat \pi)
-\frac{3}{8 \Lambda^6}\frac{1+5 \alpha_3+6\alpha_3^2}{\varepsilon^2} \L^{(4)}_{\rm Gal}(\hat \pi)+\frac{\hat \pi}{2\sqrt{\varepsilon} \mpl} \mathcal{T}\,,
\ea
where we keep the coupling to matter arbitrary in this case ($\mathcal{T}$ can correspond to the trace of the energy-momentum tensor or could include the divergence of that one). The presence of any source of matter $\mathcal{T}\ne 0$ provides a non-trivial background for $\hat \pi$. The canonically normalized fluctuations $\chi$ on top of this background, defined symbolically as $\hat \pi=\hat \pi_0(x) +\chi/\sqrt{Z}$ with $Z\sim 1+ \frac{\p^2 \pi_0 }{\Lambda^3 \varepsilon^{3/2}}+ (\frac{\p^2 \pi_0 }{\Lambda^3 \varepsilon})^2$, see the following linearized action
\ba
\L^{(\chi)} = -\frac34 (\p \chi)^2 +\frac{\chi}{2\sqrt{\e Z}}\delta \mathcal{T}\,.
\ea
In the limit $\varepsilon \to 0$, the coupling to matter thus behaves as $\varepsilon^{3/2} \chi  \delta \mathcal{T}$ and any fluctuations in the helicity-0 mode hence decouple from the rest of the rest of the fields (the matter fields and the helicity-2 part of the graviton). We therefore recover the standard Vainshtein mechanism whereby within the close vicinity of a source, the field interactions are large and are responsible for the screening of the field. This screening mechanism is completely similar to what happens in the massless limit of massive gravity.

The presence of this new realization of the Vainshtein mechanism, still within massive gravity but when taking the limit $m^2\to 2 H^2$ rather than the massless limit, opens the door for a new window of phenomenological opportunities. Within the context of massive gravity, this screening mechanism is usually used to put constrains on the graviton mass itself (\ie a bound on the deviation from General Relativity). Within this framework however one can see that many of the observational constraints can equivalently be read as a bound on the deviation from $m^2=2H^2$ rather than from General Relativity.

\subsubsection{PM limit}

The realization of the Vainshtein mechanism in the previous paragraph relies crucially on taking the parameters $\alpha_n$ to be different from the PM ones and only sending $m^2\to 2H^2$. An essential consequence of the fact that the $\alpha_n$ are different is that the symmetry \eqref{sym} is then broken non-linearly even when $\e=0$.  So there is no reasons why the coupling to matter should itself respect that symmetry (\ie it is fully consistent to choose a coupling for which  $\mathcal{T}\ne 0$).

If on the other hand we take the PM limit (either by setting the PM parameters first and sending $\e \to 0$ or by sending all the parameters to their PM values simultaneously), the mere existence of the symmetry  \eqref{sym}  in the PM case (and its non-linear realization), also fixes the coupling to matter and imposes the constraint $\mathcal{T}=0$ (which for instance can be read as a failure of conservation of energy as in \eqref{Matter-constraint}). In this case, the helicity-0 completely decouples  as soon as the constraint $\mathcal{T}$ is imposed, and there is therefore no vDVZ discontinuity to talk about.

To understand this point better, let us make an analogy with the GR limit of massive gravity. In GR, coordinate reparameterization invariance imposes the matter Lagrangian to be covariant. At the linearized level, this fixes the coupling to matter to be of the form $h\mn T^{\mu\nu}$ where $T^{\mu\nu}$ is the conserved stress-energy tensor for external matter. Since this symmetry is restored when  considering the massless limit of massive gravity, the same coupling to matter is also considered there. The only difference between the GR and the PM limit of massive gravity is that in the GR case, the symmetry only imposes $\partial_{\mu} T^{\mu \nu}=0$, but does not forbid the coupling $\pi T$ present in $h\mn T^{\mu\nu}$, so that a Vainshtein mechanism is required to efficiently decouple the helicity-0 mode. In the PM case, however, the symmetry \eqref{sym} imposes a more severe constraint $\mathcal{T}=0$, effectively setting to zero the would-be interactions between the helicity-0 mode of the graviton and matter, so that the former simply decouples without the help of any strong coupling effect.

These arguments rely fundamentally on the assumption that the PM theory in four dimensions enjoys a non-linear symmetry which fully projects out the helicity-0 mode. At the level of this work, this is nothing else but an assumption which begs to be (dis)proved.

\subsection{Partially Massless Gravity in dimension different from four}

We have seen in section \ref{sec:DL} that the helicity-0 mode fully disappears from the decoupling limit in any dimension for the specific choice of parameters (\ref{beta}-\ref{alpha_n}). Therefore, if a theory of PM gravity exists beyond the decoupling limit, then it is unique and fully determined by the parameters above.

However, in three dimensions and when working beyond the decoupling limit, a few hints suggest that PM cannot survive non-linearly. Already at the cubic level, there seems to be no symmetry that would keep the action invariant as soon as the PM quadratic mass term is introduced, \cite{Zinoviev:2006im,Zinoviev:2012yb}. This suggests that the helicity-0 mode reappears non-linearly. If this is confirmed, then the helicity-0 mode would be infinitely strongly coupled and the notion of particle for that field would be ill-defined. This would then cast serious doubts on the existence of PM theory of gravity in three dimensions. The consistent cubic interactions for a partially massless spin-2 field were also recently derived in \cite{Joung:2012rv} in any dimensions larger than three, and four dimensions were shown to be special.

Furthermore it has also been established that a conformal symmetry is present in PM {\it only} in four dimensions and that the fields then propagate on the lightcone \cite{Deser:2001xr,Deser:1983mm}. This could provide a fundamental reason explaining why PM would only make sense in four dimensions.

The decoupling limit analysis performed here can of course not shed any light on whether the conformal symmetry present in four dimensions is maintained in full generality. However we emphasize that there is no obstruction in obtaining a PM theory of gravity in an arbitrary number of dimensions  within the decoupling limit. This would therefore be the first instance where this limit fails to manifest all the degrees of freedom fundamentally present in the theory.

\section{Outlook}
\label{sec:Outlook}

In this paper we have established  the framework for deriving the decoupling limit of massive gravity on a maximally symmetric reference metric. Our results are valid for any number of dimensions and are generalizable to anti de Sitter, \cite{AdS}.

The decoupling limit allows us not only to explore the phenomenology of massive gravity on de Sitter in depth and to investigate its Vainshtein mechanism, but also to uniquely identify the only potential candidate for a fully non-linear completion of partially massless gravity. In particular we have  shown that within the decoupling limit, the helicity-0 mode completely disappears for a very specific set of parameters. Well beyond being of a pure academic interest, the existence of a four-dimensional theory of massive gravity propagating only four degrees of freedom would open an entire new window of opportunity for cosmology and phenomenology. The helicity-0 mode that appears in any other theory of massive gravity gives the strongest bound on the graviton mass bounding it to be smaller than just a few times the current Hubble constant. In partially massless gravity on the other hand, the helicity-0 mode is fully absent and could well be protected by a non-linearly realized symmetry. In that case the fundamental gravity mass could be a few orders of magnitude larger, without nevertheless affecting any solar system nor astronomical tests of General Relativity.

Besides the phenomenological interests of this model, partially massless gravity brings the hope for a new symmetry which could have a significant impact as much for gravity than for cosmology, \cite{Deser:2001xr}. It is therefore of great interest to establish whether or not the theory we have identified non-linearly here keeps the same properties beyond the decoupling limit and when including the interactions with the helicity-1 fields.

\medskip

\begin{acknowledgments}

We would like to thank C.~Bachas,  C.~Charmousis, G.~Gabadadze, K.~Hinterbichler, R.~Rosen and A.~J.~Tolley for very fruitful remarks, as well as the anonymous referee for his/her useful comments. SRP would like to thank the Universit\'e de Gen\`eve for hospitality when this work was initiated. CdR would like to thank Columbia University for its hospitality during the final stages of this work. SRP was supported by the STFC grant ST/F002998/1 and the Centre for Theoretical Cosmology when most part of this work was carried out. This work was supported by French state funds managed by the ANR within the Investissements d'Avenir programme under reference ANR-11-IDEX-0004-02.

\end{acknowledgments}

\appendix

\section{\stu fields in the ``Na\"ive approach"}
\label{appendix:naive}

In this appendix we carry out the na\"ive  approach to covariantize the reference metric $\gamma_{\mu \nu}$ a little further (for simplicity, we focus on the four-dimensional case). Hence, instead of going through the higher-dimensional method to directly derive the proper form of the covariantized metric in terms of the helicity-0 and -2 modes, we start in a more pedestrian way by ``covariantizing" the de Sitter metric as
\ba
\tilde\gamma\mn=\gamma_{ab}\p_\mu \phi^a \p_\nu \phi^b\,,
\label{naive-covariantization}
\ea
where we choose the four \stu fields $\phi^a$ to be expressed in terms of $\tilde \pi$ as follows
\ba
\label{phia}
\phi^a=x^a-\gamma^{ab}\p_b \tilde \pi\,.
\ea
The covariantized metric is then expressed in terms of $\tilde \pi$ as follows
\ba
\label{gamma_naive}
\tilde \gamma\mn & = & \gamma\mn -2\gamma_{a(\mu} \p_{\nu)}(\gamma^{ab}\p_b \tilde \pi)+\gamma_{ab}\p_\mu (\gamma^{ac}\p_c \tilde \pi) \p_\nu (\gamma^{bd}\p_d \tilde \pi) \\
&\neq& \gamma\mn -2\nabla_{\mu} \nabla_{\nu} \tilde \pi+\gamma^{\alpha\beta} \nabla_{\mu} \nabla_{\alpha} \tilde \pi\nabla_{\nu} \nabla_{\beta} \tilde \pi
\ea
where $\nabla_\mu$ denotes the covariant derivative on de Sitter. Notice therefore that in this na\"ive approach we are dealing with intrinsically non-covariant objects, the sum of which conspire to transform in the correct way for $\tilde \gamma\mn$. There is of course nothing wrong with this approach, but we would set ourselves with a very tedious formulation from the very beginning. For example, continuing along that path, we would get at quadratic order
\ba
\label{StupidWay}
&& \frac{\mpl^2}{2} m^2 \sqrt{-g} \L_{\rm der}^{(2)}\(\K\) = \sqrt{-\gamma}
\Big[-\frac{m^2}{8}\(h\mn^2-h^2\) - \frac{m^2}{2} h^{\mu\nu} Y\mn \\
&& -\frac{m^2}{4}\left\{\p_\mu(\gamma^{\alpha\nu}\p_\nu \bar \pi) \p_\alpha(\gamma^{\mu\beta}\p_\beta \bar \pi)
 + \p_\mu(\gamma^{\alpha\nu}\p_\nu \bar \pi) \gamma^{\mu \beta} \gamma_{c \alpha} \partial_{\beta}(\gamma^{cd} \partial_d \bar \pi)
-2\p_\mu(\gamma^{\mu\nu}\p_\nu \bar \pi) \p_\alpha(\gamma^{\alpha\beta}\p_\beta \bar \pi)
 \right\} \Big]\,,\nn
\ea
where we used the same notation as before \ie $ h\mn = \mpl \tilde h\mn$ and $\bar \pi = \mpl \tilde \pi $, and
\ba
\label{Ymn}
Y\mn \equiv \gamma_{(\mu\alpha}\p_{\nu)}(\gamma^{\alpha\beta}\p_\beta \bar \pi)-\gamma\mn \p_\alpha (\gamma^{\alpha\beta}\p_\beta \bar \pi )
= \nabla_\mu \nabla_\nu \bar \pi-\gamma\mn \Box \bar \pi + H \dot{\bar \pi} \(-2 \gamma\mn+\delta^0_\mu\delta^0_\nu\)\,.
\ea
The appearance of the non-covariant term $\delta^0_\mu\delta^0_\nu$ in \eqref{Ymn}, together with the fact that the terms quadratic in $\bar \pi$ on the second line of \eqref{StupidWay} are equivalent, after integrations by parts, to \\
$\frac52 m^2 H^2 \sqrt{-\gamma} \gamma^{ij}\p_i \bar \pi \p_j \bar \pi$, where the sum is over the spatial indices only, show explicitly that the covariantization performed in \eqref{naive-covariantization}-\eqref{phia} (even though technically correct) was not the most convenient one.

In particular, with the present method, one sees that $\tilde h\mn$ carries part of the helicity-0 mode in a very non trivial way: the change of variable one would need to perform in order to diagonalize the helicity-0 and -2 modes would be of the schematic non-covariant form $\tilde h\mn = \bar h\mn + m^2 \bar \pi \g\mn +{\cal O}(H \dot{\bar \pi} \delta^0_\mu \delta^0_\nu)$ (to be contrasted with the covariant redefinition $h\mn= \bar h\mn + m^2 \bar \pi \g\mn$ in our higher-dimensional approach). Of course there is nothing wrong with using this approach, and one could continue similarly to all orders, and perform a more and more complicated (and less and less covariant) field redefinition at each order in the field expansion; it is however not a very physical way to proceed and we have presented a much cleaner derivation in section \ref{sec:5d}. Once again this is a simple consequence of the fact that $\bar \pi$ is not a scalar in this case and does not properly account for the dynamics of the helicity-0 mode, as explained thoroughly in Ref.~\cite{deRham:2011rn}.

\section{$d+1$-dimensional change of coordinates}
\label{sec:Coord-transformation}

From Eqs. \refeq{Z0}-\refeq{Zi}, one deduces
\bea
\frac{\partial x^i}{\partial Z^0}&=& \frac{\partial x^i}{\partial Z^4}  =- H x^i e^{H(Y-t)} \\
\frac{\partial Y}{\partial Z^i}&=& \frac{\partial t}{\partial Z^i}  =- H x^i e^{H(Y+t)} \\
\frac{\partial x^i}{\partial Z^j}&=& \delta^i _j  e^{H(Y-t)} \\
\frac{\partial y}{\partial Z^0}&=& \half e^{HY} \left( 2 {\rm sh}(Ht)+ e^{H t} H^2  x_i^2  \right) \\
\frac{\partial y}{\partial Z^d}&=& \half e^{HY} \left( -2 {\rm ch}(Ht)+ e^{H t} H^2  x_i^2  \right) \\
\frac{\partial t}{\partial Z^0}&=& \half e^{HY} \left( 2 {\rm ch}(Ht)+ e^{H t} H^2  x_i^2  \right) \\
\frac{\partial t}{\partial Z^d}&=& \half e^{HY} \left( -2 {\rm sh}(Ht)+ e^{H t} H^2  x_i^2  \right)\,.
\eea
From these expressions, together with Eq. \refeq{explicit-Piy}, it is tedious, but easy, to find the explicit components of the tensor $\tilde \g\mn$ (Eq. \refeq{def-f}) in the coordinate system $\lbrace t,x^i \rbrace$, which are the same as the components of the tensor \refeq{explicit-f}.

\section{de Sitter contributions}
\label{sec:appendix dS}

In this appendix, we concentrate on the second type of terms in \eqref{Lmass tot}, and therefore focus on the variation
\ba
\mathcal{L}_{\rm mass}^{H^2}&=&- \frac{\Lambda^{d+2}}{8}\frac{H^2}{m^2}\ \left[\frac{\delta}{\delta H^2} \left(\sqrt{-g}\, {\cal U} \right)\right]_{h\mn=0, H^2=0}\\
&=&- \frac{\Lambda^{d+2}}{8}\frac{H^2}{m^2}\ \left[\frac{\delta}{\delta H^2} \left(\sqrt{-\g}\, {\cal U}|_{h\mn=0} \right)\right]_{H^2=0}\,,
\ea
so that for this class of contributions, the metric can be taken as being purely de Sitter with no fluctuations.
As ${\cal U}$ is a (complicated) scalar function of the tensor ${\cal K}_{\mu \nu}$, the crucial step is to compute the term linear in $H^2$ in an expansion of ${\cal K}_{\mu \nu \,  \lvert h_{\mu \nu}=0 }$, \ie we want to find $\A_{\mu \nu}$ such that
\bea
{\cal K}_{\mu \nu \,  \lvert h_{\mu \nu}=0 }=\tilde \Pi_{\mu \nu} +\frac{H^2}{2} \A_{\mu \nu} +{\cal O}(H^4) \,,
\eea
If the terms in $\T_{\mu} \T_{\nu}$ were absent in $H_{\mu \nu}$, we would find ${\cal K}_{\mu \nu \,  \lvert h_{\mu \nu}=0 }=S_{\mu \nu}$, which contributes a trivial factor $\eta_{\mu \nu} (\partial \P)^2$ to $\A_{\mu \nu}$. Now the additional terms $\T_{\mu} \T_{\nu}$ lead to new contributions to $A\mn$ which take the form $\tilde \Pi^n_{(\mu \alpha} \left(\T^{\alpha} \T^{\beta}\right)_{\lvert H^2=0}  \tilde \Pi^m_{\beta  \nu )}$ with $n,m \geq 0$. Given the explicit expression \refeq{def-Y} of $\T_{\mu}$, this contribution can be as well seen as a sum of terms of the form $ (\partial \P \cdot \tilde \Pi^m )_{(\mu} ( \partial \P \cdot \tilde \Pi^n )_{\nu)}$ so $\A_{\mu \nu}$ can be written as
\bea
\A_{\mu \nu}=\eta_{\mu \nu} (\partial \P)^2+\sum_{m,n \geq 0}c_{m, n}  ( \partial \P \cdot \tilde  \Pi^m )_{(\mu} ( \partial \P \cdot \tilde \Pi^n )_{\nu)}\,.
\label{ansatz-A}
\eea
with symmetric coefficients $c_{m,n}$. To identify these coefficients, we go back to the definition $\K_{\mu \nu}$, $\K^2\mn-2 \K_{\mu \nu } +H_{\mu \nu}=0$. To leading order in $H^2$ this implies a linear equation for $\A_{\mu \nu}$ itself,
\bea
\tilde \Pi_{(\mu}^{\alpha} \A_{\alpha \nu)}-\A_{\mu \nu}+\frac{\delta}{\delta H^2}H\mn{}_{\lvert h\mn=0, H^2=0}=0\,,
\eea
with
\ba
\frac{\delta}{\delta H^2}H\mn{}_{\lvert h\mn=0, H^2=0}= (\p \P)^2 \(\eta\mn-\tilde \Pi\mn\) - \(\tilde \Pi_{\mu \alpha}\p^\alpha \P- \p_\mu \P\)\(\tilde \Pi_{\nu \beta}\p^\beta \P- \p_\nu \P\).
\ea
Plugging the ansatz \refeq{ansatz-A} for $\A\mn$ in this relation, we deduce the recursive relation
\bea
c_{(m-1, n)}-c_{m,n}-\delta_m^1 \delta_n^1+2 \delta_{(m}^1 \delta_{n)}^0-\delta_m^0 \delta_n^0 =0 \,, \quad \forall m,n \geq 0
\label{recursive}
\eea
from which all the $c_{m,n}$'s can be computed.

Now, to use this expression into the potential $\cal U$, we first point out that for $n \geq 0$:
\bea
\frac{\delta [ \K]^n}{\delta H^2}\Big|_{h_{\mu \nu}=0, H^2=0}=\frac{n}{2} [\tilde \Pi]^{n-1} [A]
\eea
and
\bea
\frac{\delta [\K^n]}{\delta H^2} \Big|_{h_{\mu \nu}=0, H^2=0}=\frac{n}{2} [A\cdot \tilde \Pi^{n-1}]
\eea
where the explicit expression of $[A \cdot \tilde \Pi^k]$ is deduced from eq.\refeq{ansatz-A} to be
\bea
[A \cdot \tilde \Pi^k]=[\tilde \Pi^k](\partial \P)^2+ \sum_{m,n \geq 0} c_{m,n} [\phi^{m+n+k}]\,, k \geq 0
\label{Apik}
\eea
and we have defined $\tilde \Pi^0_{\mu \nu}=\gamma_{\mu \nu}$ and $[\phi^k] \equiv \partial \P \cdot \tilde \Pi^k \cdot \partial \P$. Using the recursive relation \refeq{recursive}, one can show, using $c_{1,1}=-\frac12$ and $c_{0,n}=1/2^n$ for $n \geq1$, that $\sum_{m+n=k \geq 2} c_{m,n}=0$, with which \refeq{Apik} simplifies to
\bea
[A \cdot \tilde \Pi^k]=[\tilde \Pi^k](\partial \P)^2 -[\phi^k]+[\phi^{k+1}]\,.
\label{Apik2}
\eea
Putting these results together, one can finally show recursively  that, for $n \geq 1$,
\bea
 \frac{\delta}{\delta H^2}  \left(\sqrt{-\g} \, \L_{\rm der}^{(n)} (\K)_{\lvert h_{\mu \nu}=0} \right)_{H^2=0}=
 -\sum_{m=1}^{n}f^{n,m}(\P)\  \L_{\rm der}^{(n-m)} (\tilde \Pi)\,,
\eea
with $\L_{\rm der}^{(0)}=1$ and
\ba
\label{coeff}
f^{n,m}(\P)=\frac{(-1)^m}{2} \frac{n!}{(n-m)!} \([\phi^m]-[\phi^{m-1}]+(\partial \P)^2[\tilde \Pi^{m-1}]\)\,.
\ea
Notice that so far this result is completely general and independent of the number of dimensions. The dimensionality only starts appearing once we use the property $[\tilde \Pi^0]=d$ in \eqref{coeff}.

\end{document}